\documentclass{article}

\PassOptionsToPackage{numbers, compress}{natbib}
\usepackage[preprint]{neurips_2026}

\usepackage[utf8]{inputenc} 
\usepackage[T1]{fontenc}    
\usepackage{hyperref}       
\usepackage{url}            
\usepackage{booktabs}       
\usepackage{amsfonts}       
\usepackage{nicefrac}       
\usepackage{microtype}      
\usepackage{xcolor}         
\usepackage{graphicx}       
\IfFileExists{orcidlink.sty}{\usepackage{orcidlink}}{%
  \newcommand{\orcidlink}[1]{\href{https://orcid.org/#1}{\textsuperscript{\scriptsize iD}}}}

\makeatletter
\renewcommand{\@noticestring}{Machine Learning and the Physical Sciences Workshop, NeurIPS 2026.}
\makeatother

\newcommand{\ncards}{6{,}145}

\newcommand{\site}{https://ioanazelko.github.io/darkmatterfm-registry/}

\urldef{\urlrepo}\url{https://github.com/ioanazelko/darkmatterfm-registry}
\urldef{\urlsite}\url{https://ioanazelko.github.io/darkmatterfm-registry/}
\urldef{\urlschemabrowser}\url{https://ioanazelko.github.io/darkmatterfm-registry/registry/schema_browser.html}
\urldef{\urlfailuremodes}\url{https://ioanazelko.github.io/darkmatterfm-registry/registry/schema_browser.html#bsm_model/agent_failure_modes}
\urldef{\urlgaproadmap}\url{https://ioanazelko.github.io/darkmatterfm-registry/registry/gap_roadmap.html}
\urldef{\urlprojects}\url{https://ioanazelko.github.io/darkmatterfm-registry/registry/darkmatterfm_research_roadmap.html}
\urldef{\urlcards}\url{https://github.com/ioanazelko/darkmatterfm-registry/tree/main/registry/models/cards/}
\newcommand{\npapers}{18{,}871}
\newcommand{\naccepted}{4{,}922}
\newcommand{\nfamilies}{636}
\newcommand{\nenriched}{2{,}004}

\newcommand{\nsteps}{7{,}000}
\newcommand{\ngaps}{5{,}166}
\newcommand{\ngroups}{150}
\newcommand{\nprojects}{25}

\title{\textbf{DarkMatterFM-Registry: A Machine-Readable Registry of Dark-Matter
Models Extracted from the arXiv Literature}}

\author{%
  Ioana A.~Zelko\,\orcidlink{0000-0002-7588-976X}$^{\,1,2}$\thanks{Corresponding author: \texttt{ioana.zelko@utoronto.ca}}
  \And
  Jo Bovy\,\orcidlink{0000-0001-6855-442X}$^{\,2,3}$
  \And
  Nolan Koblischke\,\orcidlink{0000-0001-5396-5824}$^{\,2,3}$
  \AND
  \parbox{\textwidth}{\centering\normalfont\small
    $^{1}$Eric and Wendy Schmidt AI in Science Fellowship, University of Toronto, Toronto, ON, Canada \\
    $^{2}$David A.~Dunlap Department of Astronomy \& Astrophysics, University of Toronto, Toronto, ON, Canada \\
    $^{3}$Dunlap Institute for Astronomy \& Astrophysics, University of Toronto, Toronto, ON, Canada
}
}

\begin{document}

\maketitle

\begin{abstract}
To use a dark-matter model one needs to know what defines it, which
observables it affects, which codes and likelihoods evaluate it, and under
what assumptions a published bound applies. That information is scattered
across tens of thousands of papers. We present \textsc{DarkMatterFM-Registry}, a
machine-readable registry of dark-matter models extracted from arXiv full
texts by a staged large-language-model (LLM) pipeline: \ncards{} structured
model cards, validated against a seven-file JSON Schema suite, ranked by
expected value for observational follow-up, and enriched for the top
${\sim}2{,}000$ with observables, constraints, and theory-to-observable
routes. Schema checks pass $97.8\%$ of cards, and an LLM-judged accuracy
study on a stratified card sample finds $81\%$ of section verdicts correct,
$18\%$ partially correct, and none hallucinated.
As a first application we turn the registry into a priority-weighted
roadmap of \nprojects{} concrete projects the community could build. The
roadmap is built by judging every computational step of every enriched
model's theory-to-observable routes as available, partial, or missing, and
aggregating the resulting typed gaps weighted by model priority. We find that what
blocks the most high-priority models is not missing theory but missing
shared infrastructure: model files for existing relic-abundance solvers, a
general non-cold Boltzmann-code framework, reusable indirect-detection
likelihoods, and a Lyman-$\alpha$ forest likelihood for non-cold dark
matter (the largest gap for structure-formation probes).
The model cards, schemas, roadmap, and
ranking are released making the project reproducible and the cards available for the community.
\end{abstract}

%
%

\section{Introduction}
\label{sec:intro}

The particle nature of dark matter remains unknown despite decades of
evidence for its gravitational effects on cosmological and galactic scales
\cite{Bertone2005,Planck2018}. With no confirmed non-gravitational signal,
the theory community has produced an enormous ecosystem of candidates:
thermal relics, axions and ultralight fields, sterile neutrinos,
self-interacting dark sectors, primordial black holes, freeze-in scenarios,
and more \cite{Feng2010,Hui2017,Tulin2018,Cirelli2024}. Each carries its
own parameters, early-Universe assumptions, imprint on the growth of cosmic
structure \cite{Bullock2017,Buckley2018}, and set of applicable constraints,
so comparisons between models collapse in practice onto a few benchmark
parameterizations, typically a thermal warm-dark-matter mass, even where
that mapping misleads. This happens just as new programs begin to probe
dark-matter microphysics through structure formation
\cite{Bechtol2019,DrlicaWagner2019}.
LLMs now make it feasible to read the literature at corpus
scale and extract schema-conforming records from full texts
\cite{Dagdelen2024}, so that instead of a static review one can build a
registry: a versioned catalog in which every model is a structured
model card. Such a registry serves human search and comparison, and we see it as a
prerequisite for agentic inference over the model space.

Our contributions are (i)~a schema suite and a registry of \ncards{} model
cards extracted from \npapers{} candidate arXiv papers, with a
follow-up-value ranking and an observational enrichment layer for the top
\nenriched{} (Sec.~\ref{sec:registry}); (ii)~a capability-gap
roadmap: every step of every enriched model's theory-to-observable routes
is judged available, partial, or missing, and the \ngaps{} resulting gaps
are aggregated into a priority-weighted ranking of \nprojects{} concrete
projects the community could build (Sec.~\ref{sec:gaps}); (iii)~validation by schema checks and an accuracy study, with the
registry's limitations (Sec.~\ref{sec:validation}). The model cards, schemas, an interactive
schema browser, the roadmap, and the project ranking are publicly
available.\footnote{\label{fn:site}Landing page linking the cards,
schemas, prompts, viewers, and scripts: \urlsite}

\section{The registry}
\label{sec:registry}

\paragraph{Anatomy of a model card.}
A card should make a model actionable: it must say what the model is,
what assumptions make it valid, which observables and constraints apply,
which routes (codes, emulators, likelihoods) evaluate it, what is
missing to evaluate it, where each claim comes from, and where an
automated agent could make a scientifically dangerous mistake. Because published bounds hold only under specific assumptions, every card carries an \texttt{agent\_failure\_modes} block listing known hazards for automated reasoning, such as mapping a model with suppressed small-scale power onto a thermal warm-dark-matter mass when its transfer function (the scale-dependent suppression of density fluctuations) is oscillatory.
We designed the schema around the questions users will ask of the registry. To start, we record each model’s phenomenological signature in its own field. This allows users to look up models by their observable effects across theory families, for example to see which models can be distinguished with a given probe. Furthermore, we list the codes and likelihoods that can be used to evaluate the model and mark those that are invalid in its regime. We give missing capabilities their own entries, so users can see which tools would have to be built for each model. Finally, we keep physics content, provenance, computation, and warnings in separate parts of the card, so that each claim can be checked against its source and corrected without changes to the rest of the card.\footnote{Schema browser: \urlschemabrowser}
\paragraph{Corpus and pipeline.}
The corpus is the arXiv full-text collection of \textsc{MINT-1T}
\cite{Awadalla2024MINT1T}: the \texttt{minty-astro-ph} subset
\cite{Smith2026MintyAstroPh} ($156{,}006$ papers) and the $69{,}093$
\texttt{hep-ph} papers of \texttt{MINT-1T-ArXiv} \cite{MINT1TArXiv}
(1994--2024). Both slices passed the same keyword filter (full text
containing ``dark matter'' and ``model''). The \texttt{astro-ph}
candidates alone then passed an LLM abstract screen tuned for recall,
asking whether the paper introduces or substantially develops a
dark-matter model; this screen only reduced the cost of the larger
\texttt{astro-ph} slice, where most keyword matches use dark matter rather
than propose a model. After deduplication \npapers{} candidates entered
extraction. We extract models from the candidates of both slices with the GPT-5.5\footnote{\url{https://developers.openai.com/api/docs/models/gpt-5.5}} language model in the open-source Codex\footnote{\url{https://github.com/openai/codex}} agent harness, which reads
each full text along with the schema and decides whether the paper introduces a
dark-matter model (\naccepted{} accepted, a rejection reason recorded
otherwise), and emits one card per model variant, giving \ncards{} cards
across ${\sim}\nfamilies$ families.\footnote{\label{fn:cards}One YAML file
per card: \urlcards} Each card is then ranked by a human-written rule
that GPT-5.5 applies to the card's fields (see Table~\ref{tab:llm}). Cards whose own content says the
model is already ruled out by data are set aside; every other card is
scored for how much observational follow-up would be worth (testability,
cosmological leverage, the probes that can check it, feasibility of the
calculation, and robustness of the extracted content) rather than for how
likely the model is to be true. For the top \nenriched{} cards a second
pass re-reads the primary and related papers and fills the observational
schemas: observables per channel, constraints with their assumptions and a
robustness grade, and routes naming the solvers, simulations, emulators,
and likelihoods that evaluate the model.

LLMs are the extraction instrument of this work. The \texttt{astro-ph} abstract
screen was a single recall-oriented question posed to Gemini~3.1 Flash-Lite\footnote{\url{https://aistudio.google.com/docs/models/gemini-3.1-flash-lite}} on each
abstract (three-way answer: novel model, not novel, unsure). The extraction, ranking, and enrichment steps ran with
GPT-5.5 in the Codex agentic harness. Finally, the gap annotation of
Sec.~\ref{sec:gaps} ran with Claude (\texttt{claude-fable-5}) as a
multi-agent workflow, one agent per five cards, reading card content plus
the annotation specification and schemas (September 2026), after a 20-card
human-reviewed pilot. Table~\ref{tab:llm} (Appendix~\ref{app:pipeline})
lists the stages: the pipeline consumed ${\approx}6.4$ billion tokens. Every stage's prompts,
decision rules, and data products are released (Appendix~\ref{app:pipeline}).
The outputs were verified by schema validation of every card, the pilot reviews,
and the accuracy study of Sec.~\ref{sec:validation}.

\section{Discovering the highest-impact missing projects for dark-matter research}
\label{sec:gaps}

\begin{figure}[t]
  \centering
  \includegraphics[width=\linewidth]{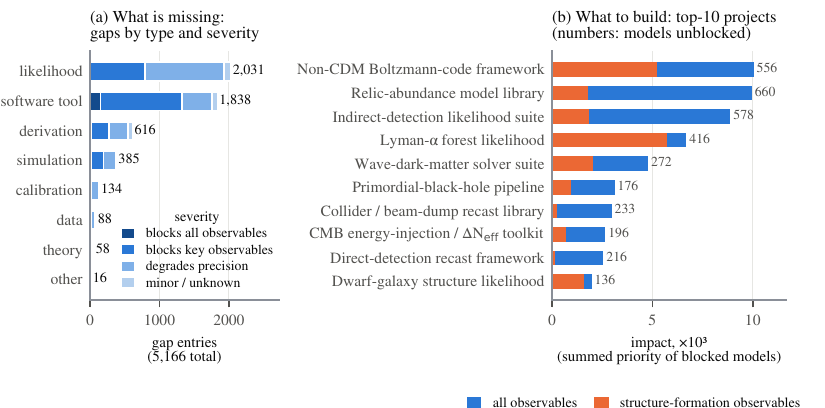}
  \caption{The capability-gap roadmap. (a)~The \ngaps{} gap entries by kind
  of missing capability and by how badly the gap blocks the model's
  evaluation; a software-tool gap is a prediction code, such as a Boltzmann
  or relic-abundance solver, that does not exist or does not cover the
  model's regime. (b)~The ten projects that unblock the most follow-up value, in
  total and through structure-formation observables alone (numbers: models
  unblocked). A Lyman-$\alpha$ forest likelihood for non-cold dark matter is
  the largest structure-formation gap.}
  \label{fig:gaps}
\end{figure}

\paragraph{Annotation.}
A model's routes record the chain of computations that connects its theory
to each observable (a Boltzmann code evolving cosmological perturbations, a
simulation or emulator for the non-linear regime, a likelihood comparing
with data). The roadmap asks the converse: which of these steps cannot
currently be done? An annotation pass assigned every route step of every
enriched card one of four statuses: available (a named public tool covers
this model's regime, or the step is simple analytic work), partial (a tool
exists but is private, needs modification, or covers a different regime),
missing, or unknown. For each missing capability it wrote a gap entry
recording its type (software tool, likelihood, derivation, simulation, and
so on), the observables and routes it blocks, a severity grade, an effort
estimate, and a concrete proposed resolution (e.g.\ ``add a decay term to
the non-cold sector of CLASS''). Nearly half of the \nsteps{} route steps
are partial and one in twelve is missing; the pass produced \ngaps{} gaps,
dominated by missing likelihoods and software tools, with missing theory
rare (Fig.~\ref{fig:gaps}a).

\paragraph{Aggregation.}
To decide which tools to build first, we rank candidate projects by the
follow-up value they unblock per unit of effort. This step is deterministic. Each gap entry is assigned a capability theme by the first match in its description or proposed resolution
from an ordered keyword list
(e.g.\ ``lyman'' for the Lyman-$\alpha$ forest
likelihood). Grouping by gap type and theme gives \ngroups{} groups, and merging each theme's groups gives one project (\nprojects{} in total); gaps that match no keyword are left out. A project's impact sums, over the distinct models it blocks, each model's follow-up priority score weighted by the severity of its worst gap in the project (1.0, 0.8, 0.4, and 0.1 for gaps that block all observables, block key observables, degrade precision, or are minor). Its cost is the mean of the effort labels that Claude gave its gaps in the annotation pass (small, medium, large, or major
program), converted to $1$, $3$, $10$, and $30$, so that each
step up is roughly three times more work (an effort index, not a time estimate).
We divide impact by cost to get the value ratio used
for the ranking, and a structure-formation view counts only the gaps that block structure-formation observables.\footnote{\label{fn:gaproadmap}Gap roadmap, with every group, project, and the
models each unblocks: \urlgaproadmap}

\paragraph{Results.}

We discovered that what keeps hundreds of dark-matter models from being tested is rarely
missing theory and usually missing shared software. Figure~\ref{fig:gaps}b and Table~\ref{tab:projects} give the ranking; all
\nprojects{} projects are online.\footnote{\label{fn:projects}Full project ranking: \urlprojects} The project with the
largest impact is a maintained Boltzmann-code framework for non-cold dark
matter \cite{Blas2011CLASS,CyrRacine2016ETHOS}: one interface for warm,
interacting, decaying, and self-interacting species. Such a code turns a
model into a prediction for cosmic structure, and $556$ models have none
that covers their regime. The best value per unit effort is a library of
model files for existing relic-abundance solvers
\cite{Belanger2018micrOMEGAs,Ambrogi2019MadDM,Bringmann2018DarkSUSY}, which
compute how much dark matter a model produces in the early Universe: the
solvers exist, but $660$ models were never written into them. For structure
formation the largest gap is a Lyman-$\alpha$ forest likelihood for non-cold
cosmologies \cite{Viel2013,Irsic2017,Murgia2017}, a comparison with the
small-scale structure seen in quasar spectra: $416$ models are constrained
only through a thermal warm-dark-matter proxy that their own cards flag as
inapplicable. Most gaps are partial rather than absent (private, unreleased,
or single-regime code), so each project is a bounded engineering task with a
named list of models it unblocks. The cards themselves invite the same use:
a theorist can pull every model predicting a given signature, an
experimentalist can check what assumptions a bound rests on, and an ML
researcher gets \ncards{} structured, source-linked records of a literature,
with the hazards for automated reasoning recorded next to the physics.

\begin{table}[t]
\caption{The eight capability projects with the highest value ratio (impact per unit effort; impact, cost, and the structure-formation (SF) view are defined in the text; cost is rounded to the nearest integer). Names are abbreviated as in Fig.~\ref{fig:gaps}; each project's full deliverable and the list of models it unblocks are listed online (footnote~\ref{fn:gaproadmap}).}
\label{tab:projects}
\centering\small
\begin{tabular}{@{}lrrrrr@{}}
\toprule
Capability project & Impact & SF & Models & Cost & Value \\
\midrule
Relic-abundance model library & 9,952 & 1,788 & 660 & 3 & 3,318 \\
Non-CDM Boltzmann-code framework & 10,048 & 5,214 & 556 & 4 & 2,716 \\
Indirect-detection likelihood suite & 8,886 & 1,824 & 578 & 4 & 2,539 \\
Lyman-$\alpha$ forest likelihood & 6,694 & 5,716 & 416 & 4 & 1,633 \\
Direct-detection recast framework & 2,561 & 167 & 216 & 2 & 1,067 \\
Collider / beam-dump recast library & 2,972 & 231 & 233 & 3 & 929 \\
Wave-dark-matter solver suite & 4,791 & 2,027 & 272 & 5 & 904 \\
CMB energy-injection / $\Delta N_\mathrm{eff}$ toolkit & 2,653 & 684 & 196 & 3 & 804 \\
\bottomrule
\end{tabular}
\end{table}

\section{Validation and limitations}
\label{sec:validation}

We validated every released card against the schema suite: $97.8\%$ of
\ncards{} cards pass, and the failures are structural anomalies queued for
re-extraction. We measured content accuracy on a random sample of $75$
cards stratified over enrichment status and taxonomy class. An LLM judge,
Claude (\texttt{claude-fable-5-1}), from a different model family than the
extractor, graded every non-empty section against the full source paper as
correct, partially correct, incorrect, hallucinated, or cannot-judge, citing
the paper location behind every non-correct verdict. Future work will include an expert re-grade of
a subsample. Of the graded verdicts,
$81\%$ were correct and $18\%$ partially correct; three were incorrect and
none hallucinated (per-section results, the grading protocol, and the
recurring defects are in Appendix~\ref{app:accuracy}; the judge's
instructions and every verdict are released with the registry). Sections recording the card's own scope (assumptions,
failure modes, identity) score above $90\%$, the physics-content sections
fall in between, and the enriched observables and constraints blocks score
lowest, near $50\%$. Two defects account for most of this loss. First, the enrichment step was instructed to draw on related papers but often attributed what it found there to the source paper; the judge, which reads only the source paper, grades such content partially correct (the same content marked as inferred is graded correct). Second, under-extraction: a value given in the paper, such as a benchmark mass, was left as unknown on the card.

\paragraph{Conclusion.}
The registry answers a question no single review can: which tools, if
built next, would let the most high-priority dark-matter models be tested.
The
\ncards{} cards are public (footnote~\ref{fn:cards}) and support other
uses, from literature search and novelty checks to structured training
data; we intend them as the knowledge backbone of agentic inference systems
that must decide which analyses apply to a model and where they do not.

\begin{ack}
We thank the authors of \textsc{MINT-1T} and of \texttt{minty-astro-ph} for
releasing the arXiv full-text corpora this work is built on. IZ is funded by the Eric and Wendy Schmidt AI in Science Fellowship through the Schmidt Sciences Foundation. JB and NK acknowledge support from the Natural Sciences and Engineering Research Council of Canada (grant RGPIN-2026-04837). This research
has made use of NASA's Astrophysics Data System and of arXiv. This work was enabled by credits provided through OpenAI's ChatGPT Futures award. This work also used API credits provided by Google through its Gemini Academic Program.
\end{ack}

\bibliographystyle{plainnat}
{\small\bibliography{refs}}

@article{Planck2018,
  author  = {{Planck Collaboration} and Aghanim, N. and others},
  title   = {Planck 2018 results. {VI}. Cosmological parameters},
  journal = {Astronomy \& Astrophysics},
  volume  = {641},
  pages   = {A6},
  year    = {2020},
  doi     = {10.1051/0004-6361/201833910},
  eprint  = {1807.06209}
}

@article{Bertone2005,
  author  = {Bertone, Gianfranco and Hooper, Dan and Silk, Joseph},
  title   = {Particle dark matter: Evidence, candidates and constraints},
  journal = {Physics Reports},
  volume  = {405},
  pages   = {279--390},
  year    = {2005},
  doi     = {10.1016/j.physrep.2004.08.031}
}

@article{Feng2010,
  author  = {Feng, Jonathan L.},
  title   = {Dark Matter Candidates from Particle Physics and Methods of Detection},
  journal = {Annual Review of Astronomy and Astrophysics},
  volume  = {48},
  pages   = {495--545},
  year    = {2010},
  doi     = {10.1146/annurev-astro-082708-101659}
}

@article{Hui2017,
  author  = {Hui, Lam and Ostriker, Jeremiah P. and Tremaine, Scott and Witten, Edward},
  title   = {Ultralight scalars as cosmological dark matter},
  journal = {Physical Review D},
  volume  = {95},
  pages   = {043541},
  year    = {2017},
  doi     = {10.1103/PhysRevD.95.043541}
}

@article{Tulin2018,
  author  = {Tulin, Sean and Yu, Hai-Bo},
  title   = {Dark Matter Self-interactions and Small Scale Structure},
  journal = {Physics Reports},
  volume  = {730},
  pages   = {1--57},
  year    = {2018},
  doi     = {10.1016/j.physrep.2017.11.004}
}

@article{Cirelli2024,
  author  = {Cirelli, Marco and Strumia, Alessandro and Zupan, Jure},
  title   = {Dark Matter},
  journal = {arXiv e-prints},
  year    = {2024},
  eprint  = {2406.01705},
  archivePrefix = {arXiv},
  note    = {arXiv:2406.01705}
}

@article{Bullock2017,
  author  = {Bullock, James S. and Boylan-Kolchin, Michael},
  title   = {Small-Scale Challenges to the {$\Lambda$CDM} Paradigm},
  journal = {Annual Review of Astronomy and Astrophysics},
  volume  = {55},
  pages   = {343--387},
  year    = {2017},
  doi     = {10.1146/annurev-astro-091916-055313}
}

@article{Buckley2018,
  author  = {Buckley, Matthew R. and Peter, Annika H. G.},
  title   = {Gravitational probes of dark matter physics},
  journal = {Physics Reports},
  volume  = {761},
  pages   = {1--60},
  year    = {2018},
  doi     = {10.1016/j.physrep.2018.07.003}
}

@article{Bechtol2019,
  author  = {Bechtol, Keith and Drlica-Wagner, Alex and Abazajian, Kevork N. and others},
  title   = {Dark Matter Science in the Era of {LSST}},
  journal = {arXiv e-prints},
  year    = {2019},
  eprint  = {1903.04425},
  archivePrefix = {arXiv},
  note    = {arXiv:1903.04425}
}

@article{DrlicaWagner2019,
  author  = {Drlica-Wagner, Alex and Mao, Yao-Yuan and Adhikari, Susmita and others},
  title   = {Probing the Fundamental Nature of Dark Matter with the Large Synoptic Survey Telescope},
  journal = {arXiv e-prints},
  year    = {2019},
  eprint  = {1902.01055},
  archivePrefix = {arXiv},
  note    = {arXiv:1902.01055}
}

@article{Dagdelen2024,
  author  = {Dagdelen, John and Dunn, Alexander and Lee, Sanghoon and Walker, Nicholas and Rosen, Andrew S. and Ceder, Gerbrand and Persson, Kristin A. and Jain, Anubhav},
  title   = {Structured information extraction from scientific text with large language models},
  journal = {Nature Communications},
  volume  = {15},
  pages   = {1418},
  year    = {2024},
  doi     = {10.1038/s41467-024-45563-x}
}

@article{Awadalla2024MINT1T,
  author  = {Awadalla, Anas and Xue, Le and Lo, Oscar and Shu, Manli and Lee, Hannah and Guha, Etash Kumar and Jordan, Matt and Shen, Sheng and Awadalla, Mohamed and Savarese, Silvio and Xiong, Caiming and Xu, Ran and Choi, Yejin and Schmidt, Ludwig},
  title   = {{MINT-1T}: Scaling Open-Source Multimodal Data by 10x: A Multimodal Dataset with One Trillion Tokens},
  journal = {arXiv e-prints},
  year    = {2024},
  eprint  = {2406.11271},
  archivePrefix = {arXiv},
  note    = {arXiv:2406.11271}
}

@misc{MINT1TArXiv,
  author       = {{mlfoundations}},
  title        = {{MINT-1T-ArXiv}: the {arXiv} subset of {MINT-1T} ({WebDataset} shards)},
  year         = {2024},
  howpublished = {Hugging Face dataset, \url{https://huggingface.co/datasets/mlfoundations/MINT-1T-ArXiv}},
  note         = {CC-BY-4.0; accessed 2026}
}

@misc{Smith2026MintyAstroPh,
  author       = {Smith, Mike},
  title        = {minty-astro-ph: the astro-ph subset of {MINT-1T-ArXiv}},
  year         = {2026},
  howpublished = {Hugging Face dataset, \url{https://huggingface.co/datasets/Smith42/minty-astro-ph}},
  note         = {CC-BY-SA-4.0; accessed 2026}
}

@article{Blas2011CLASS,
  author  = {Blas, Diego and Lesgourgues, Julien and Tram, Thomas},
  title   = {The Cosmic Linear Anisotropy Solving System ({CLASS}) {II}: Approximation schemes},
  journal = {Journal of Cosmology and Astroparticle Physics},
  volume  = {07},
  pages   = {034},
  year    = {2011},
  doi     = {10.1088/1475-7516/2011/07/034}
}

@article{CyrRacine2016ETHOS,
  author  = {Cyr-Racine, Francis-Yan and Sigurdson, Kris and Zavala, Jes{\'u}s and Bringmann, Torsten and Vogelsberger, Mark and Pfrommer, Christoph},
  title   = {{ETHOS}---an effective theory of structure formation: From dark particle physics to the matter distribution of the Universe},
  journal = {Physical Review D},
  volume  = {93},
  pages   = {123527},
  year    = {2016},
  doi     = {10.1103/PhysRevD.93.123527}
}

@article{Belanger2018micrOMEGAs,
  author  = {B{\'e}langer, Genevi{\`e}ve and Boudjema, Fawzi and Goudelis, Andreas and Pukhov, Alexander and Zaldivar, Bryan},
  title   = {{micrOMEGAs5.0}: Freeze-in},
  journal = {Computer Physics Communications},
  volume  = {231},
  pages   = {173--186},
  year    = {2018},
  doi     = {10.1016/j.cpc.2018.04.027}
}

@article{Ambrogi2019MadDM,
  author  = {Ambrogi, Federico and Arina, Chiara and Backovi{\'c}, Mihailo and Heisig, Jan and Maltoni, Fabio and Mantani, Luca and Mattelaer, Olivier and Mohlabeng, Gopolang},
  title   = {{MadDM} v.3.0: a Comprehensive Tool for Dark Matter Studies},
  journal = {Physics of the Dark Universe},
  volume  = {24},
  pages   = {100249},
  year    = {2019},
  doi     = {10.1016/j.dark.2018.11.009}
}

@article{Bringmann2018DarkSUSY,
  author  = {Bringmann, Torsten and Edsj{\"o}, Joakim and Gondolo, Paolo and Ullio, Piero and Bergstr{\"o}m, Lars},
  title   = {{DarkSUSY} 6: An Advanced Tool to Compute Dark Matter Properties Numerically},
  journal = {Journal of Cosmology and Astroparticle Physics},
  volume  = {07},
  pages   = {033},
  year    = {2018},
  doi     = {10.1088/1475-7516/2018/07/033}
}

@article{Viel2013,
  author  = {Viel, Matteo and Becker, George D. and Bolton, James S. and Haehnelt, Martin G.},
  title   = {Warm dark matter as a solution to the small scale crisis: New constraints from high redshift {Lyman-$\alpha$} forest data},
  journal = {Physical Review D},
  volume  = {88},
  pages   = {043502},
  year    = {2013},
  doi     = {10.1103/PhysRevD.88.043502}
}

@article{Irsic2017,
  author  = {Ir{\v s}i{\v c}, Vid and Viel, Matteo and Haehnelt, Martin G. and others},
  title   = {New Constraints on the free-streaming of warm dark matter from intermediate and small scale {Lyman-$\alpha$} forest data},
  journal = {Physical Review D},
  volume  = {96},
  pages   = {023522},
  year    = {2017},
  doi     = {10.1103/PhysRevD.96.023522}
}

@article{Murgia2017,
  author  = {Murgia, Riccardo and Merle, Alexander and Viel, Matteo and Totzauer, Maximilian and Schneider, Aurel},
  title   = {``Non-cold'' dark matter at small scales: a general approach},
  journal = {Journal of Cosmology and Astroparticle Physics},
  volume  = {11},
  pages   = {046},
  year    = {2017},
  doi     = {10.1088/1475-7516/2017/11/046}
}

\appendix
\section{Pipeline overview and LLM usage}
\label{app:pipeline}

\begin{figure}[h]
  \centering
  \includegraphics[width=\linewidth]{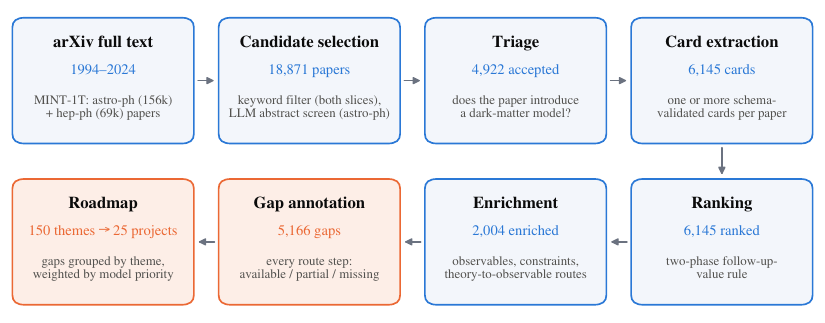}
  \caption{Construction of the registry and of the capability-gap roadmap.
  Every stage emits a versioned data product; the two orange stages are the
  application demonstrated in this paper. Counts are those of the release
  snapshot.}
  \label{fig:pipeline}
\end{figure}

Figure~\ref{fig:pipeline} summarizes the construction of the registry
(Sec.~\ref{sec:registry}) and of the capability-gap roadmap
(Sec.~\ref{sec:gaps}), and Table~\ref{tab:llm} lists the LLM stages with
the model, harness, input, and token consumption of each.

\begin{table}[h]
\centering
\caption{LLM stages of the pipeline and their token consumption. Tokens are
input plus output, with cached-prefix reads counted; GPT-5.5 stages were
$91\%$ cached input, Claude stages $92\%$. Orchestration covers the sessions
that launched, resumed, and validated the runs and built the roadmap. At
each stage we used the strongest model then available from whichever
provider we had compute resources on; no stage was chosen by comparing
models.}
\label{tab:llm}
\footnotesize
\setlength{\tabcolsep}{4pt}
\begin{tabular}{@{}lllr@{}}
\toprule
Stage & Model (harness) & Input per unit of work & Tokens \\
\midrule
Abstract screen (\texttt{astro-ph} only) & Gemini 3.1 Flash-Lite & paper abstract & $0.50$~B \\
Triage and card extraction & GPT-5.5 (Codex) & full paper text + schema prefix & $3.81$~B \\
Ranking and enrichment (top \nenriched{}) & GPT-5.5 (Codex) & card + primary and related papers & $1.53$~B \\
Capability-gap annotation & Claude (\texttt{claude-fable-5}) & card + annotation spec + schemas & $0.47$~B \\
Orchestration and validation & Claude (\texttt{claude-fable-5.1}) & --- & $0.09$~B \\
\midrule
Total & & & ${\approx}6.4$~B \\
\bottomrule
\end{tabular}
\end{table}

\paragraph{Released prompts and decision rules.}
Every LLM stage's instructions are released with the registry
(footnote~\ref{fn:site}), so that each decision summarized in
Table~\ref{tab:llm} can be inspected and re-run. Paths are relative to the
repository root.
\begin{itemize}\raggedright
  \item \emph{Abstract screen:} \texttt{registry/gemini\_screening\_prompt.md} and
    \texttt{registry/gemini\_screening\_response\_schema.json}, the
    recall-oriented question and the three-label structured-output schema
    (no rationale or confidence is recorded).
  \item \emph{Triage and extraction:} \texttt{registry/codex\_prompts/}, the
    orchestrator and worker prompts of the \texttt{astro-ph} and
    \texttt{hep-ph} runs, the output guide that defines the accept/reject
    rule, the required card fields and the provenance values, and the
    run-local schema profile the workers read.
  \item \emph{Ranking:} \texttt{registry/ranking\_rule.md}, the human-written
    two-phase rule (an elimination gate for models their own card says are
    ruled out, then scoring on about 30 card fields) that GPT-5.5 applied to
    every card.
  \item \emph{Enrichment:} \texttt{registry/codex\_prompts/enrichment\_worker.md},
    its output schema, and the policy stating which sources (the primary
    paper, ADS-linked related papers, a bounded web fallback) may support a
    structured claim.
  \item \emph{Capability-gap annotation:} \texttt{registry/capability\_gap\_prompt.md},
    the specification of how a route step is judged available, partial,
    missing, or unknown, and how gap entries are typed, graded, and resolved.
  \item \emph{Aggregation:} \texttt{registry/build\_gap\_roadmap.py} and
    \texttt{registry/build\_research\_project\_ranking.py}, the deterministic
    grouping and weighting of Sec.~\ref{sec:gaps}.
  \item \emph{Accuracy study:} \texttt{registry/accuracy\_study/}, the sample
    manifest, the judge's grading instructions, every verdict with its note,
    the browser interface for the human re-grade, and the analysis script
    (Appendix~\ref{app:accuracy}).
\end{itemize}

\section{Extraction-accuracy study}
\label{app:accuracy}

\paragraph{Sample and protocol.}
We drew a stratified random sample of $75$ cards (seed fixed and released):
$50$ enriched and $25$ base cards, covering all $23$ taxonomy classes and
the ranking tiers, and including $5$ of the schema-invalid cards as
deliberately hard cases. Every non-empty section of each card was graded
against the full text of its source paper with one of five verdicts:
correct (every substantive claim supported), partially correct (gist right,
at least one detail wrong, missing, or over-stated), incorrect (a central
claim contradicts the paper), hallucinated (specific content with no basis
in the paper), or cannot-judge. The protocol requires a cited paper location
for every verdict other than correct. For enriched sections, generic domain
knowledge presented as paper-derived is graded partially correct; the same
content marked on the card as inferred is graded correct.

\paragraph{The LLM judge.}
The grading reported here was performed by Claude
(\texttt{claude-fable-5-1}), a different model family from the GPT-5.5
extractor, as $15$ independent grader instances of $5$ cards each. Each
instance was instructed to act as a skeptical referee whose default verdict
is not ``correct'', to read the paper text in full before grading, and to
back every non-correct verdict with a quote or a pointer to the paper. The
instructions and all $1{,}359$ section verdicts with their notes are
released in \texttt{registry/accuracy\_study/}.

\paragraph{Results.}
Over $1{,}326$ graded section verdicts ($33$ further sections were
cannot-judge), $81.4\%$ were correct and $90.6\%$ correct or partially
correct; $3$ sections were incorrect and none hallucinated. Base cards score
$83.6\%$ strict ($n=385$ sections) and enriched cards $80.6\%$ ($n=941$).
Table~\ref{tab:accuracy} gives the per-section breakdown. Grader notes
identify six recurring defects, in decreasing frequency: (i)~provenance
bleed in the enriched sections, where observables and constraints attach
references or numbers from later or related papers that the source paper
never cites, presented as paper-derived (the dominant reason those sections
score low; under a stricter reading some of these attributions are
hallucinated provenance even though no content was invented);
(ii)~under-extraction, values the paper states left as unknown (mixing
angles, benchmark masses, code names); (iii)~transcription errors in
formulas and exponents, some traceable to the PDF-to-text conversion;
(iv)~parameter symbols renamed relative to the paper; (v)~misclassified
\texttt{theory\_level}; and (vi)~the three incorrect verdicts, all
directional or status errors (``excluded'' for ``highly constrained'', a
nonthermal label on a thermal relic, an inverted lifetime bound).

\begin{table}[h]
\centering
\caption{LLM-judged extraction accuracy per card section on the $75$-card
stratified sample. $n$ is the number of cards in which the section is
non-empty; ``Correct'' is the strict fraction with a Wilson $95\%$ interval;
no section was graded hallucinated. Sections present in fewer than $10$
cards are omitted.}
\label{tab:accuracy}
\small
\begin{tabular}{lrrr}
\toprule
Section & $n$ & Correct & $95\%$ CI \\
\midrule
\texttt{identity} & 75 & 92\% & [84,96]\% \\
\texttt{theory} & 75 & 79\% & [68,86]\% \\
\texttt{particle\_content} & 75 & 76\% & [65,84]\% \\
\texttt{symmetries} & 75 & 87\% & [77,93]\% \\
\texttt{interactions} & 75 & 72\% & [61,81]\% \\
\texttt{parameters} & 75 & 59\% & [47,69]\% \\
\texttt{production} & 75 & 89\% & [80,94]\% \\
\texttt{relic\_density} & 75 & 84\% & [74,91]\% \\
\texttt{cosmological\_history} & 75 & 77\% & [67,85]\% \\
\texttt{validity\_domain} & 75 & 96\% & [89,99]\% \\
\texttt{assumptions} & 75 & 100\% & [95,100]\% \\
\texttt{robustness} & 75 & 88\% & [79,94]\% \\
\texttt{agent\_failure\_modes} & 75 & 93\% & [85,97]\% \\
\texttt{phase\_space} & 73 & 88\% & [78,93]\% \\
\texttt{structure\_formation} & 64 & 88\% & [77,94]\% \\
\texttt{associated\_code} & 55 & 82\% & [70,90]\% \\
\texttt{observables} & 50 & 46\% & [33,60]\% \\
\texttt{constraints} & 50 & 50\% & [37,63]\% \\
\texttt{theory\_to\_observable\_routing} & 50 & 82\% & [69,90]\% \\
\bottomrule
\end{tabular}
\end{table}

\end{document}